\documentclass[conference]{IEEEtran}
\IEEEoverridecommandlockouts
\usepackage{cite}
\usepackage{amsmath,amssymb,amsfonts}
\usepackage{graphicx}
\usepackage{textcomp}
\usepackage{xcolor}
\usepackage{hyperref}
\usepackage{url}
\usepackage{subfigure}
\def\BibTeX{{\rm B\kern-.05em{\sc i\kern-.025em b}\kern-.08em
    T\kern-.1667em\lower.7ex\hbox{E}\kern-.125emX}}

\title{YouTube UGC Dataset for Video Compression Research}
\author{\IEEEauthorblockN{Yilin Wang, Sasi Inguva, Balu Adsumilli}
  \IEEEauthorblockA{
    \textit{Google Inc.} \\
    Mountain View, CA, USA \\
    \{yilin,isasi,badsumilli\}@google.com
  }
}

\begin{document}

\IEEEpubid{\makebox[\columnwidth]{978-1-7281-1817-8/19/\$31.00~\copyright~2019 IEEE \hfill} \hspace{\columnsep}\makebox[\columnwidth]{ }}
%
\maketitle

\begin{abstract}
Non-professional video, commonly known as User Generated Content (UGC) has become very popular in today’s video sharing applications. However, traditional metrics used in compression and quality assessment, like BD-Rate and PSNR, are designed for pristine originals. Thus, their accuracy drops significantly when being applied on non-pristine originals (the majority of UGC). Understanding difficulties for compression and quality assessment in the scenario of UGC is important, but there are few public UGC datasets available for research. This paper introduces a large scale UGC dataset (1500 20 sec video clips) sampled from millions of YouTube videos. The dataset covers popular categories like Gaming, Sports, and new features like High Dynamic Range (HDR). Besides a novel sampling method based on features extracted from encoding, challenges for UGC compression and quality evaluation are also discussed. Shortcomings of traditional reference-based metrics on UGC are addressed. We demonstrate a promising way to evaluate UGC quality by no-reference objective quality metrics, and evaluate the current dataset with three no-reference metrics (Noise, Banding, and SLEEQ).
\end{abstract}
\section{Introduction}\label{sec:Introduction}
Video makes up the majority of today’s Internet traffic. Consequently, this motivates video service companies (e.g. YouTube) to spend substantial effort to control bandwidth usage~\cite{Chen18RQO}. The main remedy deployed is typically video bitrate reduction. However, aggressive bitrate reduction may hurt perceptual visual quality at the same time as both creators and viewers have increasing expectations for streaming media quality. The evolution of new codec technology (HEVC, VP9, AV1) continues to address this bitrate/quality tradeoff. 

To measure the quality degradation, numerous quality metrics have been proposed in the last few decades. Some reference-based metrics (like PSNR, SSIM \cite{Wang04SSIM} and VMAF~\cite{Li16VMAF}) have been widely used in the industry.
A common assumption held by most video quality and compression research is that the original video is perfect, and any operation on the original (processing, compression etc.) makes it worse. Most research measures how good the resulting video is by comparing it to the original. However, such an assumption does not hold for most of User Generated Content (UGC) due to the following reasons:

\begin{itemize}
\item \textbf{Non-pristine original} \hspace{1mm} When there are visual artifacts present in the original, it is not clear whether an encoder should be putting in efforts to accurately represent those artifacts. It is necessary to consider the effect that the encoding has on those undesired artifacts, but it is also necessary to consider the effect that the artifacts have on the ability to encode the video effectively.

\item \textbf{Mismatched absolute and reference quality} \hspace{1mm} Using the original as a reference does not always make sense when the original isn’t perfect. Quality improvement may be affected by pre/post processing before/after transcoding, but reference-based metrics (e.g. PSNR, SSIM) cannot fairly evaluate the impact of these tools in a compression chain.
\end{itemize}


We created this large scale UGC dataset in order to encourage and facilitate research that considers the practical and realistic needs of video compression and quality assessment in video processing infrastructure.

A major contribution of this work is the analysis of the enormous content in YouTube in a way that illustrates the breadth of visual quality in media worldwide. That analysis leads to the creation of a statistically representative set that is more amenable to academic research and computational resources. We built a large scale UGC dataset(Section~\ref{sec:YouTube UGC Dataset Overview}), and propose (Section~\ref{sec:How the Dataset was Built}) a novel sampling scheme based on features extracted from encoding logs, which achieves high coverage over millions of YouTube videos. Shortcomings of traditional reference-based metrics on UGC are discussed (Section \ref{sec:Challenges on UGC quality assessment}), and we evaluate the dataset with three no-reference metrics: Noise~\cite{Chen16Noise}, Banding~\cite{Wang16Banding}, and Self-reference based LEarning-free Evaluator of Quality (SLEEQ)~\cite{Ghadiyaram17SLEEQ}.

The dataset can be previewed and downloaded from \href{<url>}{\textit{https://media.withyoutube.com/ugc-dataset}}.

\section{Related Work} \label{sec:Related Work}
Some large-scale datasets have already been released for UGC videos, like YouTube-8M~\cite{Haija2016YouTube8M} and AVA~\cite{Gu2018AVA}. However, they only provide extracted features instead of raw pixel data, which makes them less useful for compression research.

Xiph.org Video Test Media~\cite{XiphOrgDataset} is a popular dataset for video compression and it contains around 120 individual video clips (including both pristine and UGC samples). These videos have various resolutions (e.g. SD, HD, and 4K) and multiple content categories (e.g. movie and gaming).

LIVE datasets~\cite{LIVEVideoset,LIVENetflixDataset,LIVEMobileDataset} are also quite popular. All of them contain less than 30 individual pristine clips, along with about 150 distorted versions. The goal of the LIVE datasets is subjective quality assessment. Each video clip in the dataset was assessed by 35 to 55 human subjects.

VideoSet~\cite{VideoSet16} contains 220 5 sec clips extracted from 11 videos. The target here is also quality assessment, but instead of providing Mean Opinion Score (MOS) like the LIVE datasets, it provides the first three Just-Noticeable-Difference (JND) scores collected from more than 30 subjects.

Crowdsourced Video Quality Dataset~\cite{Sinno12018CrowdsourcedQualityDataset} contains 585 10 sec video clips, captured by 80 inexpert videographers. The dataset has 18 different resolutions and a wide range of quality owing to the intrinsic nature of real-world distortions. Subjective opinions were collected from thousands of participants using Amazon Mechanical Turk.

KoNViD-1k~\cite{konvid1k} is another large scale dataset which contains 1200 clips with corresponding subjective scores. All videos are in landscape layout and have resolution higher than $960\times{540}$. They started from a collection of 150K videos, grouping them by multiple attributes like blur, colorfulness etc. The final set was created by a ``fair-sampling'' strategy. The subjective mean opinion scores were gathered through crowdsourcing.

\section{YouTube UGC Dataset Overview} \label{sec:YouTube UGC Dataset Overview}
\begin{figure}
\centering
\includegraphics[trim={2cm 5.5cm 0cm 2cm}, clip, width=0.46\textwidth]{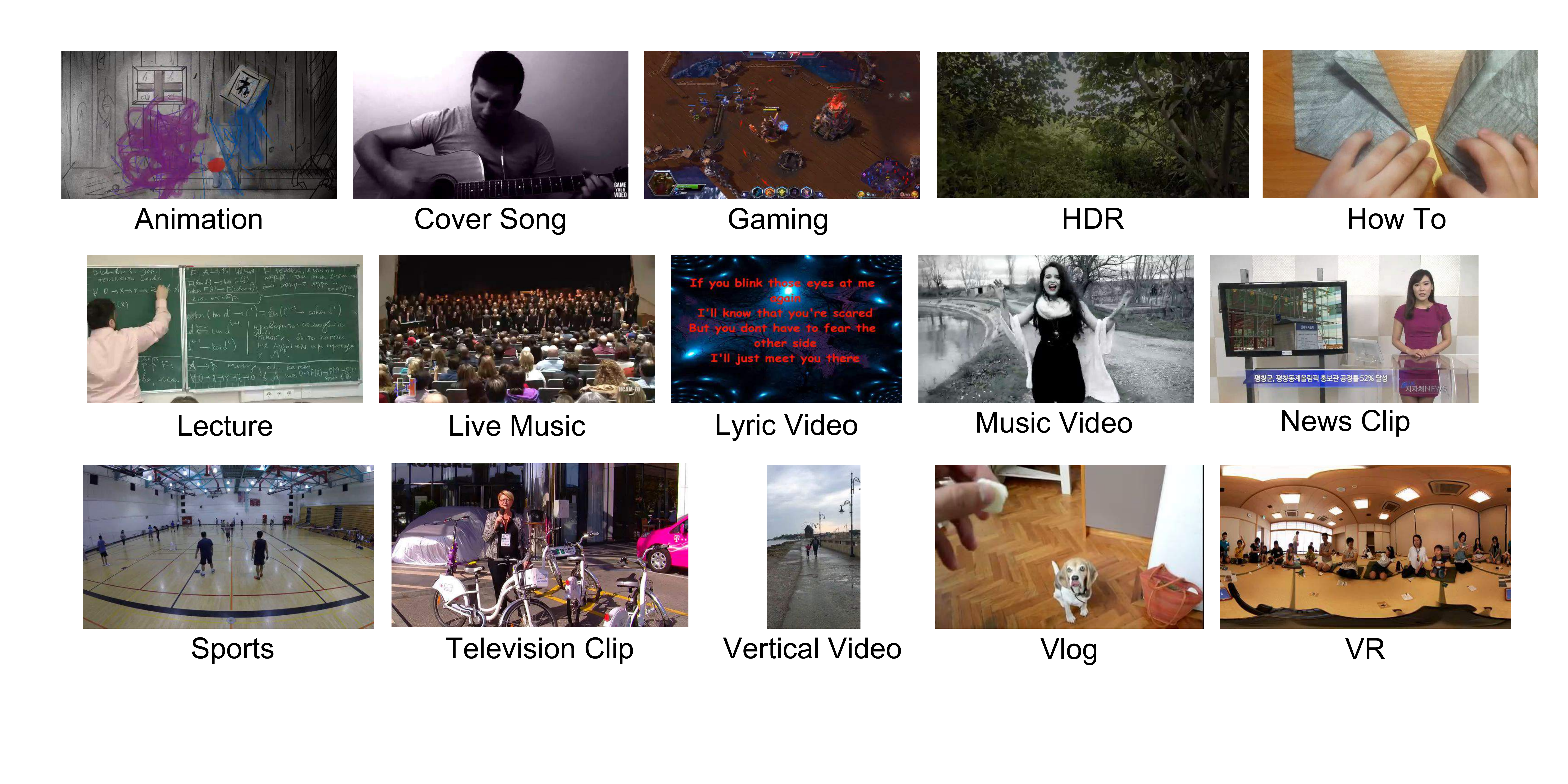}
\caption{All categories in the YouTube UGC dataset.}
\label{fig:ugc_category}
\end{figure}

Our dataset is sampled from YouTube videos with the Creative Commons license. We selected an initial set of 1.5 millions videos belonging to 15 categories annotated by Knowledge Graph~\cite{Singhal12KnowledgeGraph} (as shown in Fig.~ \ref{fig:ugc_category}): Animation, Cover Song, Gaming, HDR, How To, Lecture, Live Music, Lyric Video, Music Video, News Clip, Sports, Television Clip, Vertical Video, Vlog, and VR. The video category is an important feature of our dataset, which allows users to easily explore characteristics of different kinds of videos. For example, Gaming videos usually contain lots of fast motion, while many Lyric videos have still backgrounds. Compression algorithms can be optimized in different ways based on such category information.

Videos within each category are further divided into subgroups based on their resolutions. Resolution is an important feature revealing the diversity of user preferences, as well as the different behaviors of various devices and platforms. So it would be helpful to treat resolution as an independent dimension. In our dataset, we provided $360$P, $480$P, $720$P, $1080$P for all categories (except for HDR and VR) and $4$K for HDR, Gaming, Sports, Vertical Video, Vlog, and VR.

The final dataset contains 1500 video clips, each of 20 sec duration. All clips are in Raw YUV 4:2:0 format with constant framerate. Details of the sampling scheme are discussed in the next section.

\section{Dataset Sampling} \label{sec:How the Dataset was Built}
Selecting representative samples from millions of videos is challenging.
Not only is the sheer number of videos a challenge to process and generate features, but also the long duration of some videos (which can be in hours) makes it that much more difficult.
Compared with another large scale dataset, Konvid-1k, this set has videos sampled from a collection that is 10 times larger (1.5M vs. 150K). A video in our dataset can be sampled at any time offset from its original content, instead of taking only the first 30 sec of the original like Konvid-1k. This makes the search space for our dataset 200 times bigger (an average video being 600s long). Due to this huge search space, computing third party metrics (like Blur metric used in Konvid-1k) is resource heavy or even infeasible.

Large scale video compression/transcoding pipelines typically divide long videos into chunks and encode them in parallel. Maintaining the quality consistency among chunk boundaries becomes an issue in practice. Thus, besides the three common attributes (spatial, temporal, and color) suggested in~\cite{Winkler12AnalysisPublicDataset}, we propose the variation of complexity across the video as the fourth attribute that reflects inter-chunk quality consistency. We made the length of video clips in our dataset 20 sec, which is long enough to involve multiple complexity variations. These 20 sec clips could be taken from any segment of a video. For the 5 million hours of videos therefore, there were 1.8 billion putative 20 sec clips.

\begin{figure}
\centering
\subfigure[Spatial = 0.08]{
\includegraphics[trim={0cm 3.5cm 0cm 0cm}, clip, width=0.2\textwidth]{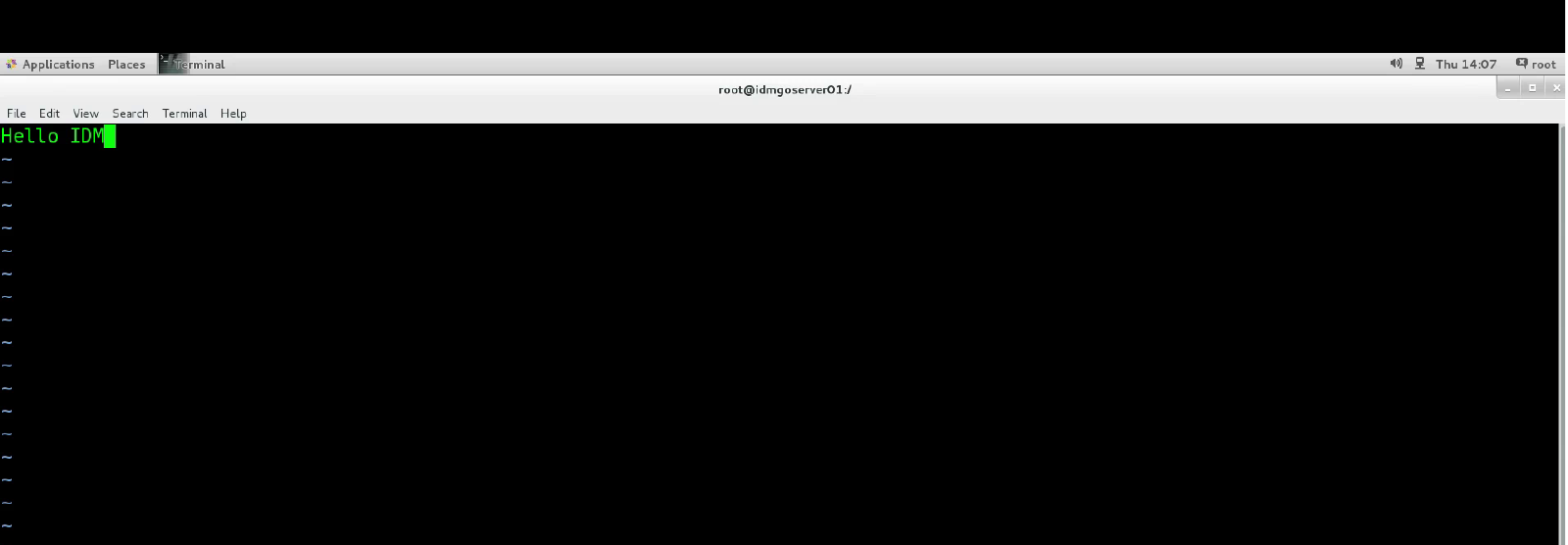}}%
\quad
\subfigure[Spatial = 4.17]{
\includegraphics[trim={0cm 3.5cm 0cm 0cm}, clip, width=0.2\textwidth]{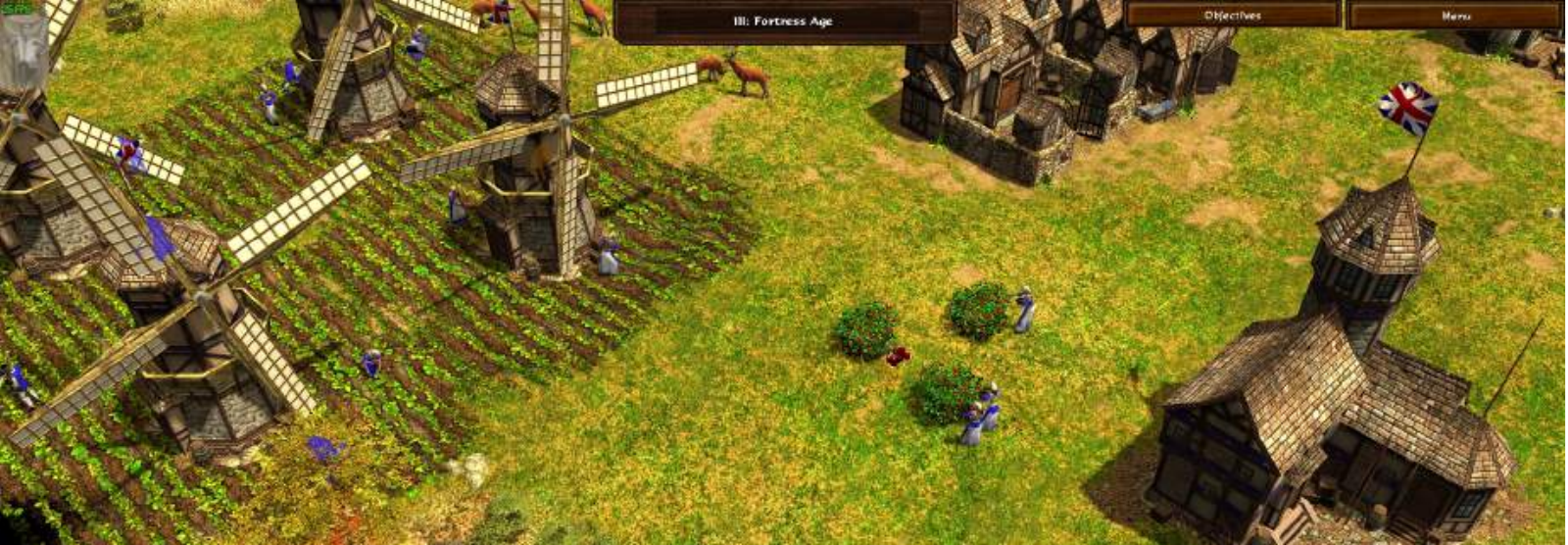}}%
\\
\subfigure[Color = 0.00]{
\includegraphics[trim={0cm 3.5cm 0cm 0cm}, clip, width=0.2\textwidth]{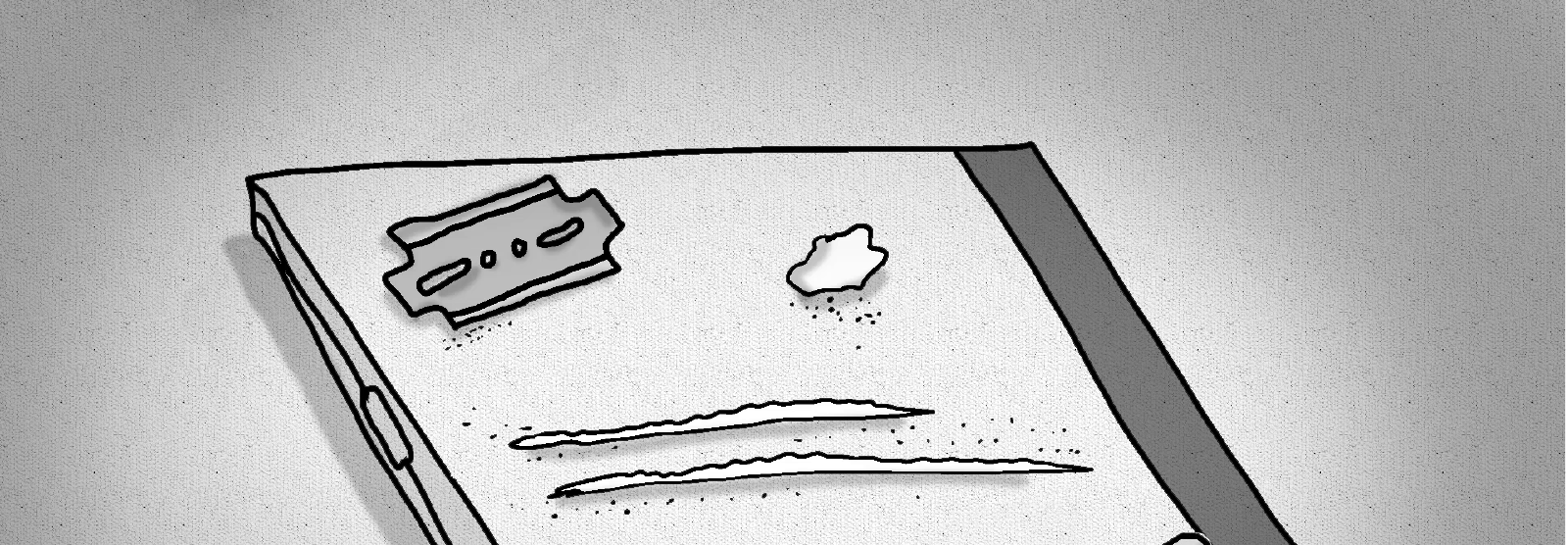}}%
\quad
\subfigure[Color = 1.07]{
\includegraphics[trim={0cm 3.5cm 0cm 0cm}, clip, width=0.2\textwidth]{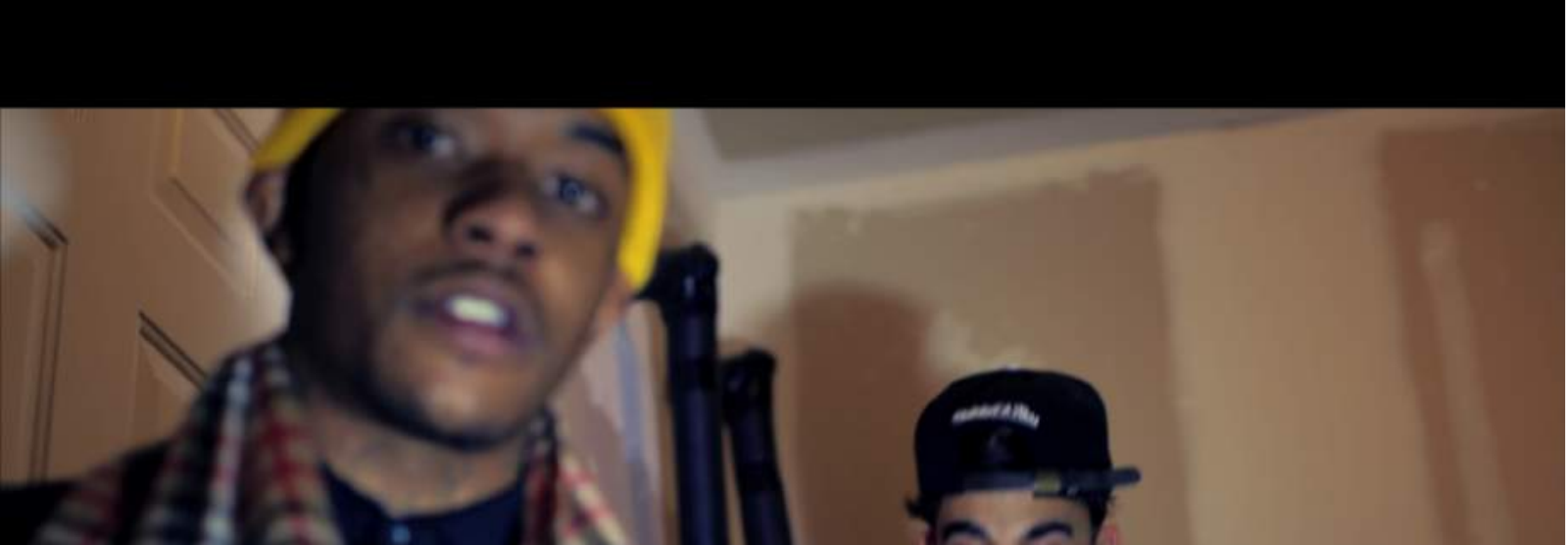}}%
\caption{Various complexity in spatial and color spaces.}
\label{fig:complexity_comparison_frame}
\end{figure}

\begin{figure}
\centering
\subfigure[Temporal = 0.07]{
\includegraphics[trim={0cm 7.5cm 0cm 0cm}, clip, width=0.2\textwidth]{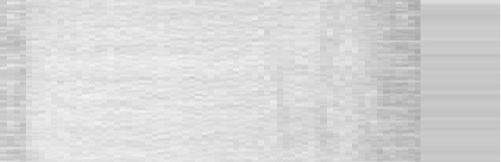}}%
\quad
\subfigure[Temporal = 1.03]{
\includegraphics[trim={0cm 7.5cm 0cm 0cm}, clip, width=0.2\textwidth]{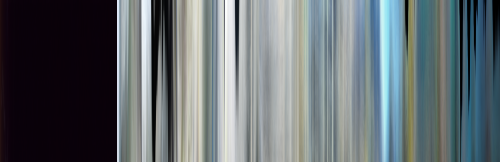}}%
\\
\subfigure[Chunk Variation = 0.05]{
\includegraphics[trim={0cm 7.5cm 0cm 0cm}, clip, width=0.2\textwidth]{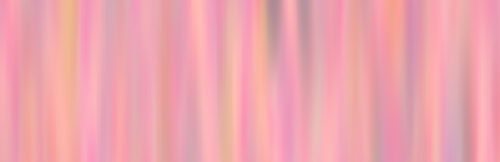}}%
\quad
\subfigure[Chunk Variation = 22.26]{
\includegraphics[trim={0cm 0cm 0cm 7.5cm}, clip, width=0.2\textwidth]{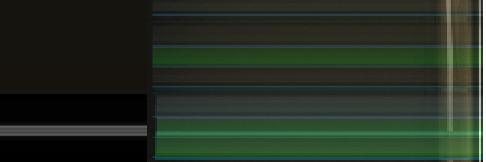}}%
\caption{Various complexity in temporal and chunk variation spaces, where the $i$th column is the row sum of the $i$th frame.}
\label{fig:complexity_comparison}
\end{figure}

We used Google's Borg system~\cite{Verma15Borg} to encode each video in the initial collection with FFmpeg H.264 encoder with PSNR stats enabled. The detailed compression settings we used are constant QP 20, fixed GOP size 14 frames with no B frames. Other reasonable settings will also work and bring similar features. The encoder reports on statistics from processing on a per frame basis. That diagnostic output was used to collect the following features over 20 sec clip stepped by 1 sec throughout each video:
\begin{itemize}
\item \textbf{Spatial} In general spatial detail in a frame is correlated with the bits used to encode that frame, when encoded as an Intra frame. Over a 20 sec chunk therefore, we calculate our spatial complexity feature as the average $I$ frame bitrate normalized by the frame area. Fig.~\ref{fig:complexity_comparison_frame}~(a) and Fig.~\ref{fig:complexity_comparison_frame}~(b) are  frames for low and high spatial complexity respectively.

\item \textbf{Color} We define color complexity metric as the ratio between the average of mean Sum of Squared Error (SSE) in U and V channels to the mean SSE in Y channel (obtained from PSNR measurements). A high score means complex color variations in the frame (Fig.~\ref{fig:complexity_comparison_frame}~(d)), and a low score usually means a gray image with only luminance changes (Fig.~\ref{fig:complexity_comparison_frame}~(c)).

\item \textbf{Temporal} The number of bits used to encode a $P$ frame is proportional to the temporal complexity of a sequence. However visual material with high spatial complexity (large $I$ frames) tends to have large $P$ frames because small motion compensation errors lead to large residuals. To decouple this effect, we normalize the $P$ frame bits by taking a ratio with the $I$ frame bits, as a fair indicator of temporal complexity. 
Fig.~\ref{fig:complexity_comparison}.(a) and Fig.~\ref{fig:complexity_comparison}.(b) show row sum map of videos with low and high temporal complexity, where the $i$th column is the row sum of the $i$th frame. Frequent column color changes in Fig.~\ref{fig:complexity_comparison}.(b) imply fast motion among frames, while small changes in temporal domain leads to homogeneous regions in Fig.~\ref{fig:complexity_comparison}.(a).

\item \textbf{Chunk Variation} To explore quality variation within the video, we measured the standard deviation of compressed bitrates among all 1 sec chunks (also normalized by the frame size). If the scene is static or has a smooth motion, the chunk variation score should be close to 0, and the row sum map has no sudden changes (Fig.~\ref{fig:complexity_comparison}~(c)). Multiple scene changes within a video (common in Gaming videos) will lead to multiple different regions on the row sum map (Fig.~\ref{fig:complexity_comparison}~(d)).
\end{itemize}

\begin{figure}
\centering
\includegraphics[trim={1.5cm 8.5cm 1.5cm 8.5cm}, clip, width=0.48\textwidth]{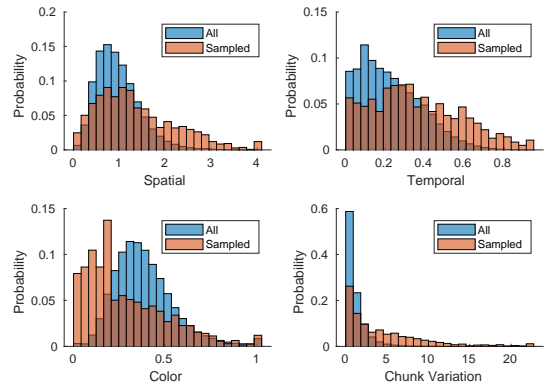}
\caption{Complexity distributions (normalized) for initial and sampled datasets.}
\label{fig:feature_space_coverage}
\end{figure}

Each original video and time offset combination forms a candidate clip for the dataset. We sample from this sparse 4-D feature space as described below:
\begin{enumerate}
\item Normalize the feature space $F$ by subtracting $\min(F)$ and dividing with $(\mathrm{percentile}(F, 99) - \min(F))$.
\item For each feature, divide the normalized range $[0, 1]$ uniformly into $N (=3)$ bins, where the last bin can go over 1 and extend up to the maximum value.
\item Permutate all non-empty bins uniformly at random.
\item For current bin, randomly select one candidate clip and add it to the dataset if and only if:
\begin{itemize}
\item The Euclidean distance of feature scores, between this clip and each of the already selected clips is greater than a certain threshold (0.3).
\item The original video that this clip belongs to, doesn't already have another clip in the dataset.
\end{itemize}
\item Move to the next bin and repeat step 4 until desired number of samples are added.
\end{enumerate}
We manually remove mislabeled clips from the 50 selected clips, leaving with 15 to 25 selected clips from each (category, resolution) subgroup. We design the sampling scheme to comprehensively sample along all four feature spaces. As the complexity distribution (normalized) shown in Fig.~\ref{fig:feature_space_coverage}, the feature scores in our sampled set are less spiky than that of the initial set. The sample distribution in color space is left skewed, because many videos with high complexity in other feature spaces are uncolorful. To evaluate the coverage of the sampled set, we divide each pairwise space into $10\times 10$ grid. The percentage of grids covered by the sampled set is shown in Table~\ref{table:Coverage rate for pairwise feature spaces}, where the average coverage rate is 89\%.

\begin{table}[htbp]
\caption{Coverage rates for pairwise feature spaces. }
\label{table:Coverage rate for pairwise feature spaces}
\centering
\begin{tabular}{|c|c|c|c|}
\hline
    & Temporal   & Color  & Chunk Variation \\
\hline
Spatial      & 93\%  & 90\% & 88\%    \\
\hline
Temporal    &  & 92\% & 88\%    \\
\hline
Color    &   &  & 83\%   \\
\hline
\end{tabular}
\end{table}


%

\section{Challenges on UGC quality assessment} \label{sec:Challenges on UGC quality assessment}
\begin{figure}
\centering
\subfigure[Original (SLEEQ = 0.21)]{
\includegraphics[trim={1cm 0cm 0cm 1cm}, clip, width=0.21\textwidth]{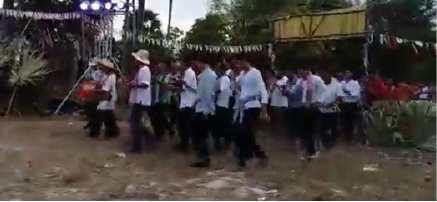}}%
\quad
\subfigure[Compressed (SLEEQ = 0.18)]{
\includegraphics[trim={1cm 0cm 0cm 1cm}, clip, width=0.21\textwidth]{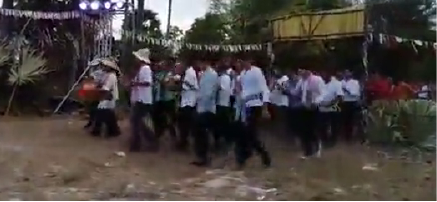}}%
\caption{Misleading reference quality scores, where PSNR, SSIM, and VMAF are 29.02, 0.86, and 58.15.}
\label{fig:bad_original_sleeq}
\end{figure}

\begin{figure}
\centering
\subfigure[Original (noise = 0.32)]{
\includegraphics[trim={4cm 2cm 4cm 2.7cm}, clip, width=0.21\textwidth]{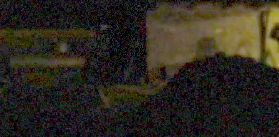}}%
\quad
\subfigure[Compressed (noise = 0.11)]{
\includegraphics[trim={4cm 2cm 4cm 2.7cm}, clip, width=0.21\textwidth]{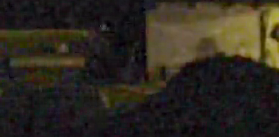}}%
\caption{Misleading reference scores caused by noise artifacts, where PSNR, SSIM, and VMAF are 30.28, 0.63, and 31.43.}
\label{fig:bad_original_noise}
\end{figure}

As mentioned in Section~\ref{sec:Introduction}, most UGC videos are non-pristine, which may confuse traditional reference-based quality metrics. Fig.~\ref{fig:bad_original_sleeq} and Fig.~\ref{fig:bad_original_noise} show two clips (id: Vlog\_720P-343d and Vlog\_720P-670d) from our UGC dataset, as well as their compressed versions (by H.264 with CRF 32). We can see that the corresponding PSNR, SSIM, and VMAF scores are bad, however the original and compressed versions are similar visually. Both the original and compressed versions in Fig.~\ref{fig:bad_original_sleeq} contain significant artifacts, while the compressed version in Fig.~\ref{fig:bad_original_noise} actually has less noise artifacts than the original one. The low reference quality scores are mainly caused by not correctly accounting for artifacts in originals.

Although no existing quality metric can perfectly evaluate quality degradation between the original and compressed versions, a possible way is to identify quality issues separately. For example, since we can tell the major quality issue for videos in Fig.~\ref{fig:bad_original_sleeq} is distortions in natural scene, we can compute SLEEQ~\cite{Ghadiyaram17SLEEQ} scores on the original and compressed versions independently, and use their differences to evaluate the quality degradation. In this case, SLEEQ scores for the original and compressed versions are 0.21 and 0.18, respectively, which implies that compression didn't introduce noticeable quality degradation. We can get the same conclusion for videos in Fig.~\ref{fig:bad_original_noise} by applying a noise detector~\cite{Chen16Noise}.

\begin{figure}
\centering
\includegraphics[trim={0cm 10cm 0cm 10cm}, clip, width=0.48\textwidth]{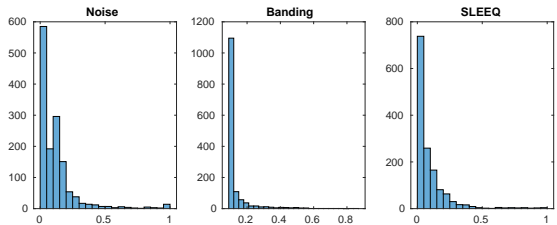}
\caption{Distribution of no-reference quality metrics.}
\label{fig:quality_distribution}
\end{figure}

\begin{figure}
\centering
\includegraphics[trim={0cm 8cm 0cm 9cm}, clip, width=0.4\textwidth]{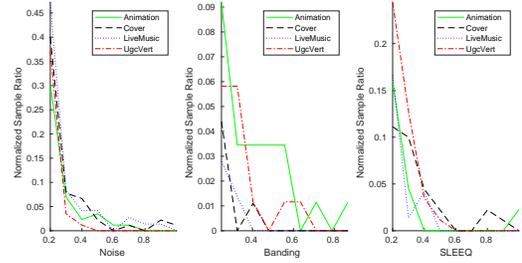}
\caption{Quality comparison for various categories.}
\label{fig:quality_metrics_comparison}
\end{figure}

We analyze perceptual quality of our UGC dataset based on existing no-reference metrics. Current evaluation includes three no-reference metrics: Noise, Banding~\cite{Wang16Banding}, and SLEEQ. The first two metrics are artifact-oriented, which can be interpreted as meaningful extents of specific compression issues. The third metric (SLEEQ) is designed to measure compression artifacts on natural scenes. All these metrics have good correlations with human ratings, and outperform other related no-reference metrics. Fig.~\ref{fig:quality_distribution} shows the distribution for the three quality metric scores (normalized to $[0, 1]$ where lower score is better) on our UGC dataset. We can see that heavy artifacts are not detected in most videos, which in some sense tells us that the overall quality of videos uploaded to YouTube is good. Fig.~\ref{fig:quality_metrics_comparison} shows some quality issues within individual categories. For example, Animation videos seem to have more banding artifacts than others, and nature scenes in Vlog tend to contain more artifacts than other categories.

%
%

\section{Conclusion} \label{sec:Conclusion}
This paper introduced a large scale dataset for UGC videos, which highly represents the videos uploaded to YouTube. A novel sampling scheme was proposed to extract features from millions of video samples, and we investigated complexity distribution as well as quality distribution for videos across 15 categories.
The difficulties of UGC quality assessment were cited. An important open question remains regarding how to evaluate quality degradation caused by compression for UGC (non-pristine reference). We hope this dataset can inspire and facilitate research on practical requirements of video compression and quality assessment.

\bibliographystyle{IEEEbib}
\bibliography{refs.bib}

\end{document}